\documentclass[sort&compress]{elsarticle}

\usepackage{wrapfig}
\usepackage{amsmath, amssymb, amsfonts}
\usepackage{eurosym}
\usepackage{graphicx}
\usepackage{tikz}
\usepackage{bbold}
\usepackage{verbatim}
\usepackage{subfigure}
\usepackage[hmargin=3cm,vmargin=3.5cm]{geometry}
\usepackage{lscape}
\usepackage[strict]{changepage}
\usepackage{units}
\usepackage{slashbox}
\usepackage{listings}
\usepackage{url}

\newcommand{\Tr}{\operatorname{Tr}}
\newcommand{\real}{\operatorname{Re}}

\newcommand{\ii}{\ensuremath{\textrm{i}}}

\newcommand{\psibar}{\bar{\psi}}

\newcommand{\mminus}{\ensuremath{D^{-1}}}

\newcommand{\loewe}{LOEWE-CSC}

\newcommand{\nsigma}{\ensuremath{\text{N}_\sigma}}
\newcommand{\ntau}{\ensuremath{\text{N}_\tau}}
\newcommand{\ndirac}{\ensuremath{\text{N}_\text{Dirac}}}
\newcommand{\nc}{\ensuremath{\text{N}_\text{C}}}
\newcommand{\nd}{\ensuremath{\text{N}_\text{D}}}
\newcommand{\vtot}{\ensuremath{\text{V}_\text{tot}}}
\newcommand{\complex}{\ensuremath{\text{C}}}
\newcommand{\dslash}{\ensuremath{{\not}D}}

\usepackage{todonotes}

\begin{document}

\title{Lattice QCD based on OpenCL}

\author[fias]{Matthias Bach}
\ead{bach@compeng.uni-frankfurt.de}

\author[fias]{Volker Lindenstruth}
\ead{voli@compeng.de}

\author[itp]{Owe Philipsen}
\ead{philipsen@th.physik.uni-frankfurt.de}

\author[itp]{Christopher Pinke}
\ead{pinke@th.physik.uni-frankfurt.de}

\address[itp]{Institut f{\"u}r Theoretische Physik, Goethe-Universit{\"a}t, Max-von-Laue-Str. 1, 60438 Frankfurt am Main}
\address[fias]{Frankfurt Institute for Advanced Studies / Institut f{\"u}r Informatik - Johann Wolfgang Goethe-Universit{\"a}t, Ruth-Moufang-Str. 1, 60438 Frankfurt am Main}

\begin{abstract}
We present an OpenCL-based Lattice QCD application using a heatbath algorithm for the pure gauge case and Wilson fermions in the twisted mass formulation.
The implementation is platform independent and can be used on AMD or NVIDIA GPUs, as well as on classical CPUs.
On the AMD Radeon HD 5870 our double precision \dslash~implementation performs at \unit{60}{GFLOPS} over a wide range of lattice sizes.
The hybrid Monte-Carlo presented reaches a speedup of four over the reference code running on a server CPU.
\end{abstract}

\maketitle

\tableofcontents


\nocite{Philipsen:2011sa}

\section{Introduction}
\label{sec:introduction}

Lattice Quantum Chromodynamics (LQCD) is the only a priory approach to describing the strong force.
State-of-the-art lattice simulations require high-performance computing and constitute actually one of the most compute intensive problems overall.
With processor clock speeds no longer improving and processors instead increasing their core count, Graphics Processing Units (GPUs) with their high peak performance and bandwidth have become an interesting platform for high-performance computing.
In June 2011 three of the top five systems in the Top500 list of supercomputers \cite{top500url} were GPU accelerated clusters.
An example of heterogeneous architecture for general purpose computing is the \loewe\ \cite{Bach2011a}, which solely consists of AMD hardware and provides two 12-core AMD Magny-Cours CPUs and one AMD Radeon HD 5870 GPU in the majority of its nodes.
Originally, it was ranked 22nd in the Top500 list of supercomputers \cite{top500url} and ranked eighth in the Green500 list of energy-efficient supercomputers (with 718 MFLOPS/Watt) \cite{greentop500url}.

Originating in the high-end computer gaming market, GPUs nowadays offer highest computing capabilities at a very attractive price-per-flop ratio.
Current high-end gaming GPUs by AMD and NVIDIA are priced at about 500 Euros.
However, while the pioneer work in the field \cite{Egri2006}, using application programming interfaces (APIs) designed for graphics rendering, was in principle platform agnostic, nearly all later developments in the field were based on NVIDIA CUDA \cite{cudaprogguide}, therefore being limited to hardware by this single GPU vendor.

We present the first application of OpenCL \cite{openclstd} to Wilson fermions, enabling the code to be run on AMD and NVIDIA GPUs and also on classical CPUs. The work, of which early prototypes have been shown in \cite{Philipsen:2011sa}, has been extended to a full hybrid Monte-Carlo application that shows major performance gains over a purely CPU based reference code.

We begin by stating the physical problem we want to solve and then explain the tools used.
Afterwards we will describe the important parts of our implementation and end with an analysis of the performance of our application.

\section{Lattice QCD and Monte Carlo simulations}
\label{sec:lqcd}

The strong interactions between elementary constituents of matter are described by Quantum Chromodynamics (QCD). In this section, we will give an introductory overview of QCD and its evaluation by means of Monte-Carlo methods. For more detailed information, we refer to \cite{degranddetar, gattringer2010quantum}.

QCD is a $SU(\nc)$ gauge theory consisting of gauge and fermion fields. An analytical access by means of perturbation theory may be applied only in regions where the coupling constant, $g$, is small. This is ensured for high momentum-transfer (``asymptotic freedom''). To explore the non-perturbative regime, euclidean spacetime is discretized on a hypercube with lattice spacing $a$. Accordingly, the QCD action $\mathcal{S}_\text{QCD}$ is replaced by a lattice version afflicted with discretization errors,
\begin{equation}
	\mathcal{S}_{\text{LQCD}} = \mathcal{S}_\text{QCD} + a\mathcal{S}_1 + a^{2}\mathcal{S}_2 + \ldots \;,
\label{lqcd}
\end{equation}
and continuum physics can be obtained in the limit $a \rightarrow 0$. 

In statistical physics, the central object is the partition function $\mathcal{Z}$ of the system, which allows for the measurement of an observable $A$ of interest. On the lattice, the expectation value of $A$ is then:
\begin{align}
	\langle A \rangle &= \mathcal{Z}^{-1} \int \mathcal{D}U \mathcal{D}\psi \mathcal{D}\psibar A   \exp \left\{ - S_\text{LQCD}[U] \right\} \nonumber \\
	&= \mathcal{Z}^{-1} \int \mathcal{D}U  A  \det D [U] \exp \left\{ - S_\text{gauge}[U] \right\}\;,
\end{align}
where the exponential in the first line is the Boltzmann factor if one identifies $S_\text{LQCD} = \beta H$. The grassmann-valued fermion fields $\psi$ can be integrated out exactly, yielding the determinant of the fermion matrix $D$, and one is left with an integral over all possible gauge configurations $U$, which are the relevant degrees of freedom. It is convenient to rewrite the fermion determinant as an integral over a bosonic field $\phi$ (\textit{pseudo fermions}), 
\begin{equation}
	\det D [U] \sim \int \mathcal{D}\phi \exp \left\{ - \phi^{\dagger} \mminus[U] \phi \right\}\;,
	\label{pseudo fermions}
\end{equation}
giving an effective action $S_\text{eff}[U, \phi] =  S_\text{gauge}[U] + \phi^{\dagger} \mminus[U] \phi$. Since it is unfeasible to solve this high-dimensional integral, importance sampling methods are used to generate an ensemble of $N$ gauge configurations $\{U_m\}$ using the Boltzmann-weight $p[U, \phi] =\exp\left\{ - S_\text{eff}[U, \phi] \right\}$ as probability measure. Then, $\langle A \rangle$ can be approximated by
\begin{equation}
	\langle A \rangle \approx \frac{1}{N} \sum_{m} A[U_m]\;.
\end{equation}

\subsection{Choice of Lattice action}

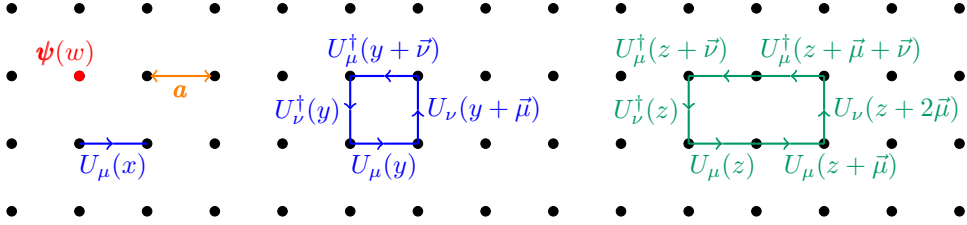
\begin{figure}
  \begin{minipage}[t]{0.95\linewidth}
 \centering
  \begin{tikzpicture}[scale=.9]
  \foreach \i in {0,...,14} \foreach \j in {0,...,3} \filldraw (\i,\j) circle (2pt);
  \draw[blue,thick,->] (1,1) -- (1.5,1);
  \draw[blue,thick] (1.5,1) -- (2,1);
  \draw[blue] (1.5,1) node [anchor=north] {$U_\mu(x)$};
  \draw[red] (.8,2) node [anchor=south] {$\pmb\psi(w)$};
  \filldraw[red] (1,2) circle (2pt);
  \draw[<->,thick,orange] (2.05,2) -- (2.95,2);
  \draw[orange] (2.5,2) node [anchor=north] {$\pmb a$};
  \draw[blue,thick,->] (5,1) -- (5.5,1);
  \draw[blue,thick] (5.5,1) -- (6,1);
  \draw[blue] (5.5,1) node [anchor=north] {$U_\mu(y)$};
  \draw[blue,thick,->] (6,1) -- (6,1.5);
  \draw[blue,thick] (6,1.5) -- (6,2);
  \draw[blue] (6,1.5) node [anchor=west] {$U_\nu(y +  \vec{\mu})$};
  \draw[blue,thick,->] (6,2) -- (5.5,2);
  \draw[blue,thick] (5.5,2) -- (5,2);
  \draw[blue] (5.5,2) node [anchor=south] {$U^{\dagger}_\mu(y + \vec{\nu})$};
  \draw[blue,thick,->] (5,2) -- (5,1.5);
  \draw[blue,thick] (5,1.5) -- (5,1);
  \draw[blue] (5,1.5) node [anchor=east] {$U^{\dagger}_\nu(y)$};
  \draw[green!60!blue,thick,->] (10,1) -- (10.5,1);
  \draw[green!60!blue,thick] (10.5,1) -- (11,1);
  \draw[green!60!blue] (10.5,1) node [anchor=north] {$U_\mu(z)$};
  \draw[green!60!blue,thick,->] (11,1) -- (11.5,1);
  \draw[green!60!blue,thick] (11.5,1) -- (12,1);
  \draw[green!60!blue] (12.25,1) node [anchor=north] {$U_\mu(z + \vec{\mu})$};
  \draw[green!60!blue,thick,->] (12,1) -- (12,1.5);
  \draw[green!60!blue,thick] (12,1.5) -- (12,2);
  \draw[green!60!blue] (12,1.5) node [anchor=west] {$U_\nu(z + 2\vec{\mu})$};
  \draw[green!60!blue,thick,->] (12,2) -- (11.5,2);
  \draw[green!60!blue,thick] (11.5,2) -- (11,2);
  \draw[green!60!blue] (12.25,2) node [anchor=south] {$U^{\dagger}_\mu(z + \vec{\mu} + \vec{\nu})$};
  \draw[green!60!blue,thick,->] (11,2) -- (10.5,2);
  \draw[green!60!blue,thick] (10.5,2) -- (10,2);
  \draw[green!60!blue] (9.75,2) node [anchor=south] {$U^{\dagger}_\mu(z + \vec{\nu})$};
  \draw[green!60!blue,thick,->] (10,2) -- (10,1.5);
  \draw[green!60!blue,thick] (10,1.5) -- (10,1);
  \draw[green!60!blue] (10,1.5) node [anchor=east] {$U^{\dagger}_\nu(z)$};
  \end{tikzpicture}

  \end{minipage}

  \caption{Sketch of lattice discretization at lattice spacing $a$. $w, x, y, z$ denote lattice sites, $\psi$ a fermion field and $U$ a gauge link, respectively. Also shown are the plaquette $P_{\mu\nu}(y)$ and the rectangle  product of links $R_{\mu\nu}(z)$.}
  \label{fig:lat_sketch}
\end{figure}

There is some freedom in the specific form of the lattice action, since it has to agree with QCD in the continuum limit only.
Different discretizations will suffer differently from discretization artifacts and often also do not hold all the continuum properties. 
Especially the discretization of the fermion matrix is non-trivial, since the naive one gives rise to non-physical fermion copies (``doublers''). 
It is also possible to add improvement terms irrelevant in the continuum to the lattice action, which help to reduce discretization errors.

At this point, it is adequate to insert some words related to the dimensions of the quantities involved and to fix our notation. 
We will denote the spatial and temporal extent of the system with \nsigma\ and \ntau, respectively, so that the total volume of the system is $\vtot = \nsigma^{3} \times \ntau$. 
The number of colors, Dirac indices and spacetime dimensions are denoted as \nc, \ndirac\ and \nd, respectively and they take on the values 3, 4 and 4. 
On each lattice site $x$, the component $\phi(x)$ of $\phi$ is a complex valued, ($\ndirac \times \nc$)-dimensional vector, whereas the gauge field $U = U_\mu(x)$ is a complex valued $\nc \times \nc$ matrix linking $x$ and its neighbour $x + \vec{\mu}$ in spacetime-direction $\vec\mu$ (therefore called links, see fig. \ref{fig:lat_sketch}).

In our numerical studies \cite{Philipsen:2008gq, Ilgenfritz:2009ns, Burger:2011zc} we employ the tree-level Symanzik improved Wilson action in the gauge sector,
\begin{equation}
	S_\text{tlsym} = \frac{\beta}{\nc} \sum_x \left( c_0  \sum_{\mu, \nu > \mu} \left\{ 1 - \real \Tr (P_{\mu\nu}(x) )\right\}  + c_1 \sum_{\mu, \nu } \left\{ 1 - \real \Tr (R_{\mu\nu}(x) )\right\}  \right)\;.
\end{equation}
Here, $P_{\mu\nu}(x)$ and $R_{\mu\nu}(x)$ denote path-ordered plaquette and rectangle products of link variables (see fig. \ref{fig:lat_sketch}) and the parameters are $\beta = 6/g^{2}$, $c_0 = 1 - 8c_1$ and $c_1 = 1/12$. 
The unimproved gauge action is regained setting $c_1$ to zero. 

In the fermionic sector we use the so-called twisted mass Wilson fermions \cite{Shindler:2007vp} to simulate two flavours of fermions, whose matrix on the lattice reads
\begin{align}
	D_\text{tm}^{\pm} &= \left( 1 \pm 2 \ii a \kappa \mu \gamma_5 \right) \delta_{xy}\delta_{\alpha\beta}\delta_{ab}  - \frac{\kappa}{2} \sum_{\mu} \left(1 - \gamma_{\pm\mu}\right)_{\alpha\beta} U_{\pm\mu}(x)_{ab} \delta_{n+\vec\mu,y}  \nonumber \\
	&= M_\text{diag}^{\pm}  + \dslash\;,
\end{align}
with shorthand notation $\gamma_{-\mu} = - \gamma_\mu$ and $U_{-\mu}(x) = U_\mu(x - \vec\mu) ^{\dagger}$. 
The $\gamma_\mu$s are matrices in Dirac space satisfying $\{\gamma_\mu, \gamma_\nu\} = 2 g_{\mu\nu}$ and $a, b, \alpha, \beta$ are color and Dirac indices, respectively. 
The sign in the diagonal mass matrix $M_\text{diag}^{\pm}$ corresponds to up and down flavour. 
Two mass parameters appear, the twisted mass $\mu$ and the usual $m$ (via the hopping parameter $\kappa = (2(a m + 4))^{-1}$). 
Pure Wilson fermions can be reobtained at vanishing twisted mass, $D_\text{Wilson} \equiv D_\text{tm}(\mu = 0)$.
Twisted mass fermions have the advantage that if $\kappa$ is tuned such that the renormalized quark-mass vanishes for pure Wilson fermions, the $\mathcal{O}(a)$ discretization errors in (\ref{lqcd}) vanish too. 
For $\mu \neq0$ the mass is then determined solely by $\mu$ and the system is said to be tuned to \textit{Maximal Twist}. 
In addition, for two flavours one has $\det D_\text{tm} = \det \left( D_\text{Wilson}^{2} + 4\kappa^2\mu^2\right)$.
Thus, the twisted mass acts as an infrared regulator, which is important in avoiding so called exceptional configurations, for which the fermion matrix has very small eigenvalues rendering its inversion ill-defined (see below).
The most used discretization type besides Wilson fermions are staggered fermions. Regarding numerics, these differ a lot from the Wilson type since the ``staggering'' essentially means that on each site there is only one spinor component instead of four. 

We now want to dwell on the role the fermion matrix plays in LQCD simulations a bit more closely. For simplicity, we will drop the indices on $D$ from now on. 
To evaluate the fermion determinant in the partition function one has to know the inverse of $D$ (see (\ref{pseudo fermions})). 
In addition, the entries of \mminus\ are related to the fermion propagator, thus its inversion is also crucial for measuring fermionic observables. 
To give an example, the two-point correlation function at sites $n$ and $m$ for a generic meson $\bar{d}(n) \Gamma u(n)$ is given by:
\begin{align}                                                                                                                                                       C_\Gamma &\equiv \langle      \bar{u}(m) \Gamma d(m) \bar{d}(n) \Gamma u(n) \rangle = - \Tr\left[ \Gamma (\mminus)(n,m)\Gamma \gamma_5 ((\mminus)(n,m))^{\dagger} \gamma_5 \right] \;.                                                                                                                                \label{propflavourmult2}                                                                                                                            \end{align}
For more involved observables, more occurrences of \mminus\ will appear.

$D$ is a ($\nc \times \ndirac \times \vtot$)-dimensional sparse square matrix, so one uses iterative Krylov space based methods to calculate \mminus\ indirectly out of equations like 
\begin{equation}
  D \phi = \psi \Rightarrow \phi = \mminus \psi \;.
  \label{solvereq}
\end{equation}
During the inversion, the matrix-vector product $D \phi$ has to be carried out many times.
The most common solvers used are CG and BiCGStab.
The convergence of the solver depends inversely on how light the simulated quarks are, and also scales with $\vtot$.
This renders the inversion the most involved part of the simulation. 
In addition, there is a lot of machinery around to simplify (\ref{solvereq}), such as mass-preconditioning and, most prominently, even-odd-preconditioning. 

In a simulation the dimensions of the lattice strongly depend on the physical problem under investigation. 
In order to rule out finite size effects the spatial volume is typically large.
For simulations of thermal systems, temperature is set by $T = 1/a\ntau$ and thus lattices have a rather small temporal extent here.
Accordingly, in vacuum simulations, the opposite is the case.
On the other hand, thermal systems require simulations over ranges of parameter values and higher statistics compared to vacuum simulations.
Typical lattices have currently $\nsigma \approx 32,\ \ntau \approx 12$ in thermal studies and $\nsigma \approx 64,\ \ntau \approx 128$ for vacuum simulations.
Therefore a thermal lattice is typically simulated using less parallelism. 

\subsection{Ensemble generation}

The standard simulation algorithm to generate QCD gauge configurations is the \textit{Hybrid Monte-Carlo (HMC)} algorithm \cite{Duane:1987de}, where the effective action is embedded in a fictitious classical system governed by the Hamiltonian $H = \frac{1}{2} P^{2} + S_\text{eff}[U, \phi]$ by introducing Gaussian momenta $P$ conjugate to $U$. 
Starting from a given configuration $(U, P)$, the system is evolved over a time $\tau$ to a new configuration $(U', P')$, according to the hamiltonian equations of motions with a force F, 
\begin{align}
\dot{P} &= - \partial S_\text{eff} / \partial U \nonumber \equiv F\;,\\
\dot{U} &= P\;.
\end{align}
This evolution is carried out by numerical integration, the most common
integration schemes being the leapfrog- and the second order minimal (2MN)
scheme. 
$\phi$ and $P$ are chosen with a Gaussian distribution initially and $\phi$ is held constant throughout the evolution of the system. 
Since the numerical integration is not exact, a metropolis step is carried out in the end, thus ensuring detailed balance. 
This means that the new configuration is only accepted with a probability $\min \left( 1, \exp(H[P', U'] - H[P, U]) \right)$.
Here, $P = P_\mu(x)$ is a real-valued, $(\nc \times \nc - 1)$ dimensional vector, as is the force $F = F_\mu(x)$.
The latter has three contributions with our choice of lattice action:
\begin{align}
	F = F^{\text{gauge}}_{\text{plaquette}} + F^{\text{gauge}}_{\text{rectangles}} + F^{\text{fermion}}\;,
\end{align}
where the gauge part of the action has been divided into plaquette and rectangle parts. 
These are proportional to sums of link products, so called staples.
For example,  $(F^{\text{gauge}}_{\text{plaquette}})_\mu(x) \sim \sum_{\mu \neq \nu} \tilde{P}_{\mu\nu}(x)$, where we denote the staple $\tilde{P}_{\mu\nu}(x)$ as the plaquette product of links without $U_\mu(x)$ (see fig. \ref{fig:lat_sketch}).
Note that these sums are no longer elements of $SU(\nc)$.

By means of the identity $\partial_U \mminus = \mminus (\partial_U D) \mminus$, the fermion force can be evaluated. 
Note that this means that one needs to invert $D$ for the input of the force calculation. 
In addition, this part of the  force receives contributions from the \dslash\ part of the fermion matrix only since our diagonal matrix does not depend on $U$.

Techniques exist to refine the integration of the equations of motions. 
$\tau$ is usually divided into smaller substeps and the inverter can be preconditioned by introducing additional pseudo fermions via $\det(D) = \det(A) \frac{\det(D)}{\det(A)}$, where $A$ denotes a fermion matrix with a different mass than $D$ (\textit{mass preconditioning}).
Furthermore, it is beneficial to integrate cheaper force contributions more often (\textit{multiple timescales}).
For details we refer to \cite{Urbach:2005ji}.

\begin{wrapfigure}{L}{0.3\textwidth}
 \centering
\begin{verbatim}
 L(0) = U
 for i < m = # subgroups:
  W = L(i) * Staple
  w = project(W,i)
  v = update(w)
  V = extend(v,i)
  L(i) = L(i-1) * V
 U = L(m)
\end{verbatim} 
 \caption{Heatbath algorithm}
        \vspace{-2mm}
 \label{fig:heatbath_sketch}
\end{wrapfigure}

In many cases, one is interested in QCD-like systems without fermions, what is then called pure gauge theory. 
In principal, this gives $\det M = 1$ in the equations above and the HMC algorithm is still applicable. 
Nevertheless, the so-called \textit{heatbath} algorithm \cite{Creutz:1980zw, Cabibbo:1982, Kennedy:1985} provides an alternative and more direct way.
It is based on an exact algorithm for $SU(2)$ that updates a specific link according to its neighbours. 
This can be extended to the general $SU(\nc)$ case by systematically reducing the $SU(\nc)$ link to a number of $SU(2)$ subgroups, which for the $\nc = 3$ case is usually also 3.
Figure \ref{fig:heatbath_sketch} shows a sketch of the algorithm, where \verb+W+, \verb+U+, \verb+V+, \verb+L+ denote $SU(\nc)$ links and \verb+w+, \verb+v+ $SU(2)$ links, respectively. 
\verb+project+ and \verb+extend+ denote reduction and extension to and from the \verb+i+th subgroup.
The \verb+update+ routine corresponds to the aforementioned exact $SU(2)$ update or to an overrelaxation update \cite{Petronzio1990}, which serves to cover the whole configuration space more quickly. 
Only in the first update random numbers are necessary.

\section{OpenCL and Graphic Cards}
\label{sec:opencl}

Table \ref{tab:cpu-gpu-theo-peak} shows an overview of available GPUs and CPUs.
One sees immediately that GPUs surpass CPUs in peak performance as well as in memory bandwidth.
However, one also notes the drop in performance when going from single to double precision on the GPU.
To take advantage of this performance an appropriate programming model and a certain understanding of the hardware architecture is required.
One of these programming models is provided by OpenCL.

\begin{table}
  \centering
  \begin{tabular}{@{} l|l|c|c|c @{}}
                          & Chip            & Peak SP  & Peak DP  & Peak BW   \\
                          &                 & [GFLOPS] & [GFLOPS] & [GB/s] \\
   \hline
   AMD Radeon HD 5870     & Cypress         & 2720     & 544      & 154       \\
   AMD FirePro V7800      & Cypress         & 2016     & 403      & 128       \\
   AMD Radeon HD 6970     & Cayman          & 2703     & 683      & 176       \\
   AMD Radeon HD 7970     & Tahiti          & 3789     & 947      & 264       \\
   AMD FirePro W8000      & Tahiti          & 3230     & 810      & 176       \\
   \hline
   NVIDIA Tesla C1060     & Tesla           & 933      & 78       & 102       \\
   NVIDIA GeForce GTX 280 & Tesla           & 933      & 78       & 142       \\
   NVIDIA GeForce GTX 480 & Fermi           & 1345     & 132      & 177       \\
   NVIDIA GeForce GTX 580 & Fermi           & 1581     & 198      & 192       \\
   NVIDIA Tesla M2090     & Fermi           & 1331     & 665      & 177       \\
   NVIDIA GeForce GTX 680 & Kepler          & 3090     & 258      & 192       \\
   \hline
   AMD Opteron 6172       & Magny-Cours     & 202      & 101      & 42.7      \\
   AMD Opteron 6278       & Interlagos      & 307      & 154      & 51.2      \\
   Intel Xeon E5-2690     & Sandy Bridge EP & 371      & 186      & 51.2      \\
  \end{tabular}
  \caption{Theoretical peak performance of current GPUs and CPUs. SP and DP denote single and double precision, respectively. BW denotes bandwidth }
  \label{tab:cpu-gpu-theo-peak}
\end{table}

\subsection{OpenCL}

GPU applications generally consist of a controlling program (``host''), running on the CPU, that executes suitable smaller programs (``kernels'') on the GPU.
All the memory handling is performed by the host program.
The most prominent library of such kind is NVIDIA's CUDA \cite{cudaprogguide}.
Virtually all existing LQCD applications are based on CUDA, at the disadvantage that these are destined to run on NVIDIA hardware exclusively \cite{Clark:2009wm, Babich:2010mu, Bonati:2011dv, Alexandru:2011ee, Cardoso:2011xu}.
\begin{wrapfigure}{L}{0.35\textwidth}
\vspace{-5pt}
 \centering
 \includegraphics[scale=.26]{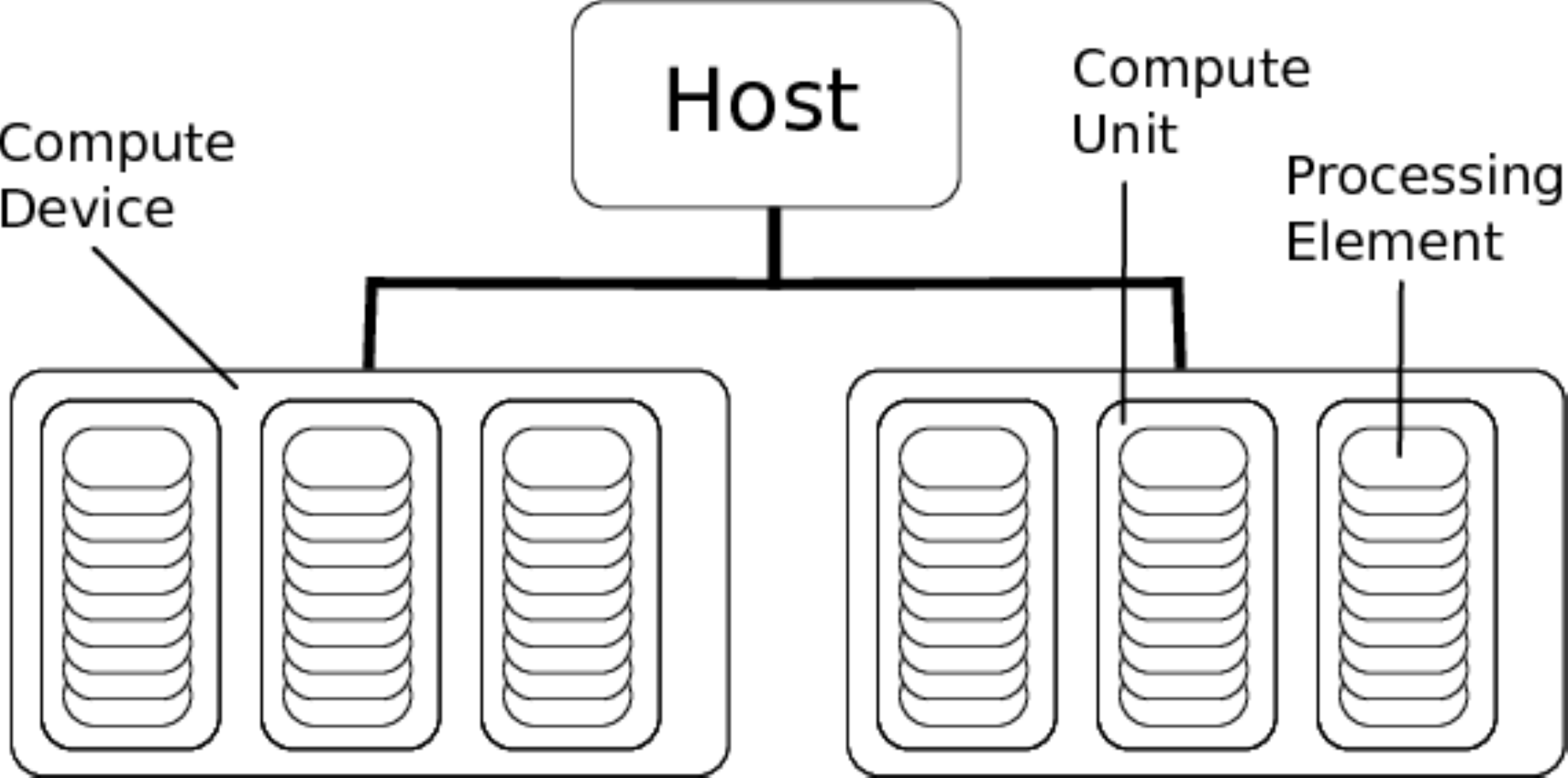}
 \caption{OpenCL Concept}
\vspace{-5pt}
 \label{opencl_concept}
\end{wrapfigure}
A hardware independent approach to parallel computing is given by the \textit{Open Computing Language} (OpenCL \cite{openclstd}), which is an open standard to perform calculations on heterogeneous computing platforms.
Thus, in OpenCL one is not determined to use CPUs or GPUs, but it is in principal possible to combine these and distribute the computations among all available computing ``devices'' (see fig. \ref{opencl_concept}).
For this OpenCL defines a programming language that is based on C99.
Implementations of OpenCL can be found both from AMD (AMD Accelerated Parallel Processing (APP) \cite{amdprogguide}, formerly ATI Stream SDK) and NVIDIA (as part of CUDA), as well as other vendors.
As LQCD applications are predominantly written in CUDA, OpenCL performance results are scarcely found.
There is but one reported for a HMC using staggered fermions \cite{Bonati:2011dv}, where a significantly lower performance of OpenCL compared to CUDA is reported.

\subsection{GPU Architecture and OpenCL Terms}

Like modern CPUs, GPUs are multi-core Single Instruction Multiple Data (SIMD) processors.
However, where CPUs are tuned towards processing each thread as fast as possible, the GPU architecture is tuned towards high throughput over thousands of threads.

While the SIMD model on CPUs is based on vector registers, where a single processor executes an instruction in parallel for multiple elements in the vector register, the GPUs implement a variant known as Single Instruction Multiple Threads (SIMT).
On the GPU registers are seen as scalar by each thread, however the execution units, called ``processing elements'' in OpenCL terminology, share a common instruction decoder.
Therefore a core, called ``compute unit'' in OpenCL, always executes a group of threads in lock-step.
The group of lock-stepped threads is basically equivalent to the SIMD thread on a CPU.
However, the memory access is more flexible due to the scalar nature of the thread execution.
On CPUs the group executed on a compute unit in lock-step might contain only one processing element.

GPUs provide a larger set of registers than CPUs.
The AMD Radeon HD 5870 provides 16384 registers, each \unit{16}{bytes} in size.
The NVIDIA GTX 480 provide 32768 registers of \unit{4}{bytes} each.
These registers are dynamically mapped to threads.
Therefore, a compute unit can run either a smaller number of threads using even more than a hundred registers each, or more than a thousand threads, each using only a dozen registers each.
Like hyper-threading on a CPU, running more threads will allow to hide memory latencies, increasing overall throughput.
The scheduling of the thread groups is performed by a hardware scheduler with minimal overhead.
Registers stay allocated to each thread from its creation until it finished execution.

The memory architecture of GPUs is more complex than that of CPUs, which appears uniform to the user.
It is split into multiple logical regions.
Global memory is the normal main memory of the GPU that can be read and written to by all threads running on the GPU.
Private memory is a part of global memory that is partitioned among all threads running on the GPU.
When addressing into private memory each thread accesses its own partition.
This memory is also used as a swapping place for registers if the register file cannot hold a threads full working set.
Registers swapped into the private memory are also known as scratch registers.
As private memory is part of the global memory it shares the same performance characteristics.
This means, large latencies can be incurred when accessing data from local memory.
Therefore, usage of scratch registers usually comes with a large performance penalty.
Another part of global memory is constant memory.
That memory can only be written to from the host.
GPUs usually are able to cache accesses to this memory and broadcast values from this memory to all threads very efficiently.
In addition modern GPUs also provide a local memory.
Just like registers, local memory is on-die and can be accessed with similar performance.
Local memory is shared between threads running on the same compute unit and can be used as a user programmed explicit cache.
Allocations of memory on the GPU are referred to as ``Buffers'' by OpenCL.

Stemming from their graphics tradition, GPUs originally only had dedicated read-only caches for constant memory and textures.
Textures are images stored in memory in a special format.
On the AMD Radeon HD 5870 and 6970, when keeping to some restrictions, the AMD OpenCL compiler is capable to automatically utilize the texture cache to access buffers which are only read by a kernel.
More modern GPUs like the NVIDIA GTX 480 and the AMD Radeon HD 7970 provide multi-level read-write caches.
While CPU caches target to minimize latencies in memory access for a single thread, GPU caches are shared by many threads.
One of their main functions is to coalesce access by multiple threads to close-by addresses into single memory transactions.

\section{Implementation strategy}
\label{sec:implemenation}

Since OpenCL needs a quite different approach to LQCD than existing software, we decided to start from scratch instead of modifying an existing application.
Consequently, all parts of the simulation code are carried out in OpenCL.
We set up the  host program in \texttt{C++}, which nicely allows for independent program parts using \texttt{C++} classes and also naturally provides extension capabilities.
The main object is the class \texttt{gaugefield}, where the initialization process of OpenCL is incorporated.
It manages the physical gauge field when several devices are used and holds an application-specific number of \texttt{opencl\_device}-
\begin{wrapfigure}{r}{0.55\textwidth}
 \centering
  \includegraphics[scale=.6]{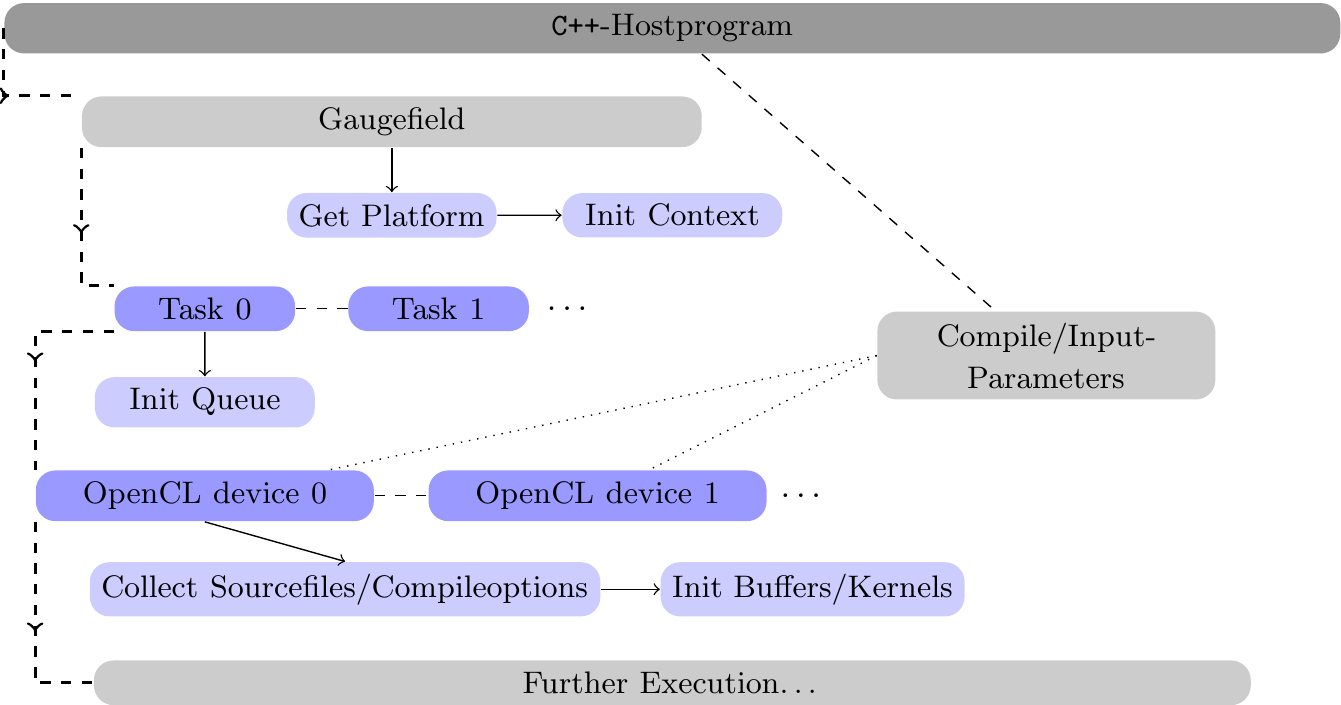}
 \caption{Schematic flow of program initialization}
        \vspace{-4mm}
 \label{prog-flow}
\end{wrapfigure}
objects.
This class in turn contains all compute device-related parts, such as kernels, and eventually executes the kernels on a specific device.
Child classes of \texttt{gaugefield} and \texttt{opencl\_module} contain problem-related functionality and provide algorithmic logic.
For hybrid applications, different ``tasks'' can be defined to carry out a physics problem. These may then again contain a number of device objects themselves.
The OpenCL environment has to be initialized before the actual calculations can be carried out (Fig. \ref{prog-flow}).
To generate the kernels, the code files are collected, compiled and linked into an OpenCL ``program`` using the OpenCL compiler.
To ease debugging we build each kernel as a stand-alone program.
The binary files, which the compiler produces during the kernel compilation, can be reused at a later point and also provide information about the kernel, e.g. register usage statistics, important for benchmarking and optimization.
Thus, the kernels are created only at runtime of the application, which allows to pass runtime parameters  (e.g. \texttt{NT}, \texttt{NS}, \ldots) to them as compile-time parameters.
We can avoid many kernel arguments like this, and parameters are ``hard coded'' into the kernel code.
On the device all data types are implemented as structures, with all the required operations defined for them.
Although this might in some cases require more registers than simply operating on arrays of scalars stored in main memory, we opted for this implementation strategy for its better code readability.

It is not trivial to estimate how much the register overhead of the structure based implementation strategy is, as the exact register requirements highly depend on the optimizer.
The possibly higher register requirements of the structure based approach can easily be seen when looking at the addition of two structures of four floats.
When operating on arrays of scalars there only has to be space for three floats in registers.
As addition is element-wise, only one float from each operand has to be loaded at the same time.
Additionally there has to be room for one element of the result.
Using actual structures four times this space is required, as the whole structure has to be stored for each operand and the result.
In the case of spinors register requirements would increase from three floats to 72.
This is, however, a worst case situation as for most operations, e.g. multiplication of a SU3 matrix with a spinor, more than one element of each structure is required at the same time anyway.
In addition, given the high latency of GPU memory it does not make sense to completely serialize handling of each element in a structure.
Therefore the register usage should be higher even when performing all operations using scalar types, as the optimizer will use different registers for different elements to enable the exploitation of instruction-level parallelism.

As up till now they have not been relevant for the overall runtime, all LAPACK operations have been implemented in a straightforward manner.
The only optimization was implicitly given by the data type storage format which is described below.
ILDG compatible I/O has been implemented  as well as the Ranlux \cite{Luescher1994100} PRNG (Pseudo Random Number Generator), as it is the standard choice for LQCD simulations.
We use the original implementation on the host while on the device we use RANLUXCL\footnote{https://bitbucket.org/ivarun/ranluxcl}, an open-source OpenCL implementation of Ranlux.
For testing purposes we have also implemented the NR3 \cite{nr3} generator, but it has not been used for the measurements in this paper.
Initialization of the random number generator follows the usual Ranlux rules which are applied across the host and the device, where each OpenCL thread runs on its own Ranlux PRNG state.

Since on different GPU drivers we have observed multiple miscompilations of our code during development, we added regression tests for most of our OpenCL functionality.
This allows us to quickly check new drivers for incorrect output.
A special challenge is that the likelihood for compiler errors scales with code complexity.
Thus, a function might work perfectly in a simple test case but will produce errors when integrated into a larger kernel.
Therefore it is important to not only test each building block for regressions, but also repeatedly check whether they still work as expected when being used in larger kernels.

\subsection{Memory requirements}

LQCD calculations are always memory bound.
To see that consider the theoretical peak performances collected in table \ref{tab:cpu-gpu-theo-peak} and the characteristics of some kernels displayed in table \ref{tab::rw_flops}.
Especially for the fermion related kernels, looking at the respective ratios of bandwidth to flops (numerical density) shows how the memory bandwidth dominates performance.
One the other hand, the heatbath related kernels may be expected to perform well on GPUs, since here the bandwidth is somewhat less important, compared to the flops the kernels perform. 
However, the numerical density is still low compared to the peak flops per bandwidth ratio of the GPUs.

Table \ref{tab::mem_req} shows that the gauge field is the biggest object to store.
In addition, one needs more than one field of each species during the simulations.
However, there are possibilities to reduce these sizes, most prominent the aforementioned even-odd preconditioning for the spinorfields.
In general, each gauge link may be displayed with $\nc^2-1$ real numbers, one for each generator of the group.
Since this representation induces additional computational overhead, other methods are more feasible.
For $\nc = 3$, one column of the matrix can be reconstructed from the other two exactly, since it has to obey $\vec c = \vec a \times  \vec b$ , so one can discard 6 floats of the full matrix (\verb+REC12+).
Additional restrictions on the left 12 numbers allow to save 2 or 4 more floats (\verb+REC10+ and \verb+REC8+, respectively).
However, the last procedure may run into arithmetic problems during the reconstruction \cite{Clark:2009wm}.
As the reconstruction adds additional complexity to the code we have not used it so far except for the \dslash\ kernel.
Most of our kernels are already at the edge of the register limit and additional code complexity tends to increase the register pressure, even if with perfect register reuse no additional registers would be required.

\subsection{Global memory storage formats}

Our first implementation of the \dslash~kernel used an array of structures (AoS) storage format for the gauge and spinor fields.
Observing this kernel reaching only about \unit{22}{GB/s} of memory throughput on the AMD Radeon HD 5870 we started studying characteristics of the memory controller.

First we checked the effects of utilizing the texture cache, which on AMD hardware can be done with minimal code modification.
While this provides a significant speedup, the achieved \unit{46}{GB/s} are still far from that GPUs theoretical bandwidth limit of \unit{155}{GB/s}.

Another speedup can be reached by specifying a proper alignment for the larger data types.
The key is to use largest possible alignment that is still a divisor of the data type size.
If no alignment is specified the compiler will use the alignment of the smallest contained data type, resulting in superfluous memory fetches.
While the number of fetches in some cases could be reduced even further by using an alignment that is larger than the data type, it turns out that the additional bandwidth required overcompensates the benefit.
Using proper alignment of all types the code reaches \unit{68}{GB/s}, which is still far from the performance limit.

\begin{figure}
\centering
\includegraphics{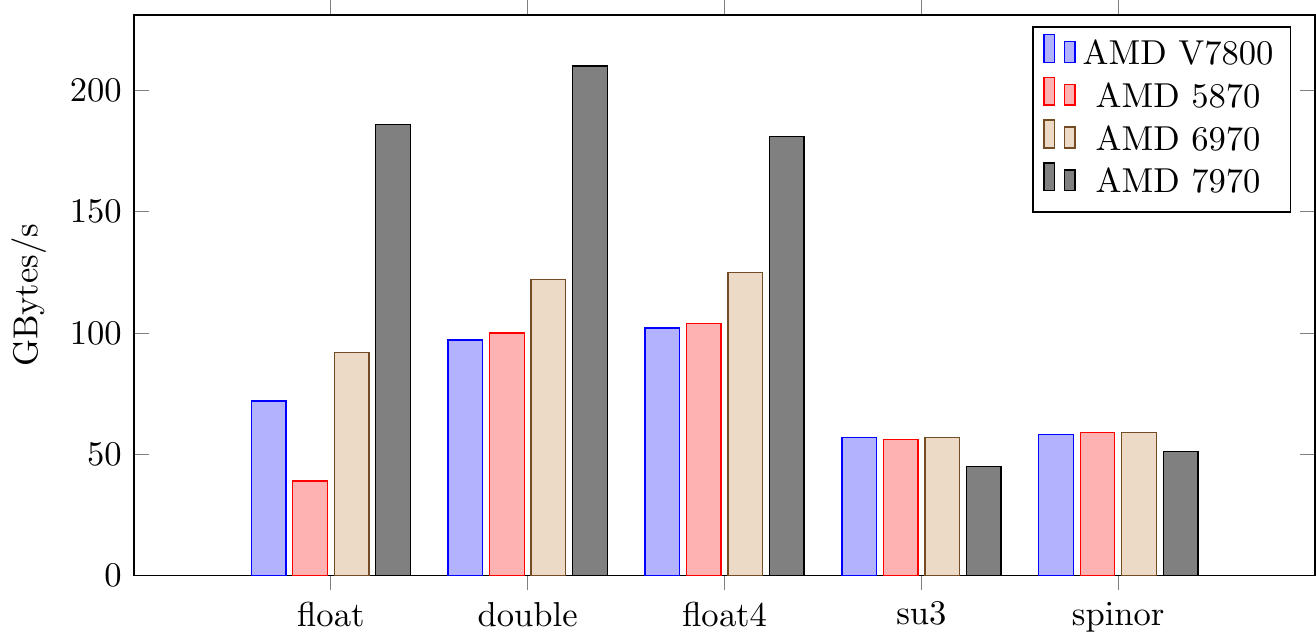}
\caption{Copy performance using different data types to copy a buffer of \unit{100}{MiB}. Note that the AMD Radeon HD 5870 was run using the Catalyst 11.7 driver, while all other measurements where performed using Catalyst 12.4.}
\label{fig:bw_copy}
\end{figure}

Figure \ref{fig:bw_copy} shows how using different data types for global memory access effects the speed of copying a fixed size buffer on the GPU.
The \lstinline|float| and \lstinline|double| types are scalar data types of \unit{4}{~bytes} and \unit{8}{~bytes} size.
The \lstinline|float4| type is a built-in data type of OpenCL that acts like a \lstinline|struct| of four floats and therefore has a size of \unit{16}{~bytes}.
The types \lstinline|su3| and \lstinline|spinor| are complex \lstinline|struct|s of size \unit{144}{~bytes} and \unit{192}{~bytes}, respectively.
The figure shows that the best way to access memory is using the \lstinline|double| or \lstinline|float4| data type.
Actually all other types with the same size and alignment will work, too.
Therefore we now use a double precision floating point complex type as the base storage type for all our data in GPU global memory.
From size and alignment this type is equivalent to the \lstinline|float4| type.
All other types are stored in a SoA fashion based on this type.
This allows to utilize memory bandwidths of up to \unit{120}{GB/s} on the AMD Radeon HD 5870.
Care has to be taken, though, as SoA increases the register pressure.
This can lead to register spilling and therefore cause a performance penalty.

Another important effect is given by the mapping of the gauge field indices to memory addresses.
As we are using even-odd preconditioning, neighboring threads will always access either links originating in even or those originating in odd sites only.
If links originating in even sites are stored interleaving with those originating in odd sites, which is the naive format, and a SoA pattern is used, a memory access will span data of both types of links.
As neighboring threads will only require one type of link at that moment, half the memory bandwidth will be wasted.
Therefore we split our gauge field such that links originating in even sites are stored separate by those originating in odd sites.

\subsection{Common code for CPUs and GPUs}

As OpenCL can be used both for CPU and GPU programming, we use a single source code for the CPU and the GPU implementation.
To cater for the different architectures, we introduces some abstractions to achieve best performance in both worlds.

An important difference between the CPU and the GPU is the optimal looping strategy.
On CPU loops perform best when each CPU works on its own consecutive block of memory.
Thereby, it can best utilize its time-local cache to reduce the number of actual requests performed on the memory.
On the GPU however, the best memory throughput is achieved if consecutive cores read consecutive elements from memory.
Therefore, a loop should always move though the data in a strided way.
We use a macro \verb+PARALLEL_FOR+ to implement these different looping strategies transparently.

Another important difference between CPU and GPU is that the GPU prefers a SoA pattern, while the CPU tends to prefer an AoS pattern.
Therefore we encapsulated all memory accesses into separate functions which transparently perform the SoA conversion if required.
When moving data onto or from a device the same automatism occurs, ensuring best memory access patterns on all devices.

While OpenCL provides vector data types, which the AMD platform uses for vectorization on the CPU, we did not use those in our code.
Besides complication of the source code, they would increase the amount of registers required on the GPU which are already a sparse resource.

\begin{table}
\begin{minipage}[t]{0.45\linewidth}
\centering
\begin{tabular}{c|c|c}
  kernel                   & RW-Size & Flops \\
  \hline
  $\dslash$                & 2880    & 1632  \\
  $F^{\text{gauge}}_{\text{plaquette}}$ & 2800    & 2717  \\
  $F^{\text{gauge}}_{\text{rectangles}}$      & 13168   & 14813 \\
  $F_{\text{fermion}}$            & 4736    & 1748  \\
  heatbath                 & 2880     & 3912  \\
  overrelax                & 2880     & 3846  \\
  saxpy                    & 608     & 64
\end{tabular}
\caption{Read-Write (RW) sizes in Bytes and flop sizes for some HMC related kernels (for double precision). All numbers are per site (and direction). ``saxpy'' corresponds to the algebraic operation $\vec z = \alpha \vec x + \vec y$. }
\label{tab::rw_flops}
\end{minipage}
\hspace{0.1\linewidth}
\begin{minipage}[t]{0.45\linewidth}
\centering
\begin{tabular}{c|c|c}
			  & general size [\vtot]                       & Bytes [\vtot] \\
  \hline
  $\phi$      & $\ndirac \times \nc \times \complex$       & $192 $        \\
  $\phi_{eo}$ & $(\ndirac \times \nc \times \complex) / 2$ & $96 $         \\
  $U$         & $\nc^{2} \times \nd \times \complex$       & $576 $        \\
  $U_{\text{REC12}}$ & $2\nc \times \nd \times \complex$          & $384 $        \\
  $U_{\text{REC10}}$ & $(2\nc-1) \times \nd \times \complex$      & $320 $        \\
  $U_{\text{REC8}}$  & $( 2\nc - 2) \times \nd \times \complex$   & $256 $        \\
  $P, F$      & $(\nc^{2} - 1 ) \times \nd $               & $256 $        \\
\end{tabular}
\caption{Overview over memory requirements of LQCD quantities in double precision per site (and direction). \complex\ denotes the size of one complex number (2 real numbers). }
\label{tab::mem_req}
\end{minipage}
\end{table}

\section{Performance results}
\label{sec:performance}

In this section we test our implementations of the heatbath and HMC algorithm and in addition report on the performance of the \dslash, the crucial time consuming part for fermionic observables and the HMC itself.
We tested our LQCD implementations on different architectures with different lattice sizes and compared the results to existing applications and literature data.
Since we are interested in thermal systems, we benchmarked our programs on lattice sizes ranging from \nsigma = 16, 24, 32, 48 and \ntau = 4, 8, 12, 16.
The \dslash\ and heatbath kernels have also been benchmarked for some additional values of \ntau\ to explore a wider range of used memory sizes.

As double precision results are desired for physical measurements we concentrated on such precision calculations throughout. However, single precision calculations do promise twice the performance for a given device's memory bandwidth and have provided good results in other applications running on the same GPUs\cite{Gerhard2012a}. Therefore our results are only a lower bound for the achievable performance. For the heatbath algorithm, where no summations over the whole lattice are required, single precision results are given.

The GPUs used are an AMD Radeon HD 5870, as it is used in the LOEWE-CSC, and an AMD FirePro V7800, which is the professional grade version of the AMD Radeon HD 5870.
It is clocked somewhat lower and equipped with more memory.
In addition we also used the newer AMD Radeon HD 6970 and the AMD Radeon HD 7970, which is the latest GPU available by AMD.
For comparison, the \dslash\ benchmarks have also been executed on the NVIDIA GeForce GTX 480, which is of comparable age to the AMD Radeon HD 5870, and the NVIDIA GeForce GTX 680, which is the latest GPU available by NVIDIA.
To show the flexibility of OpenCL, we also simulated on Intel Xeon E5520, AMD Opteron 6172 and AMD Opteron 6278 CPUs.
All of the benchmarks, except those on the AMD Radeon HD 7970, used Catalyst 12.4 which is the most current available AMD GPU driver available at the time of writing.
On the AMD Radeon HD 7970 the Catalyst 12.2 was used.
The NVIDIA GPUs were using version 295.41 of NVIDIA's GPU driver.

\subsection{Heatbath performance}

\begin{figure}[h]
\centering
\includegraphics[width=.8\textwidth]{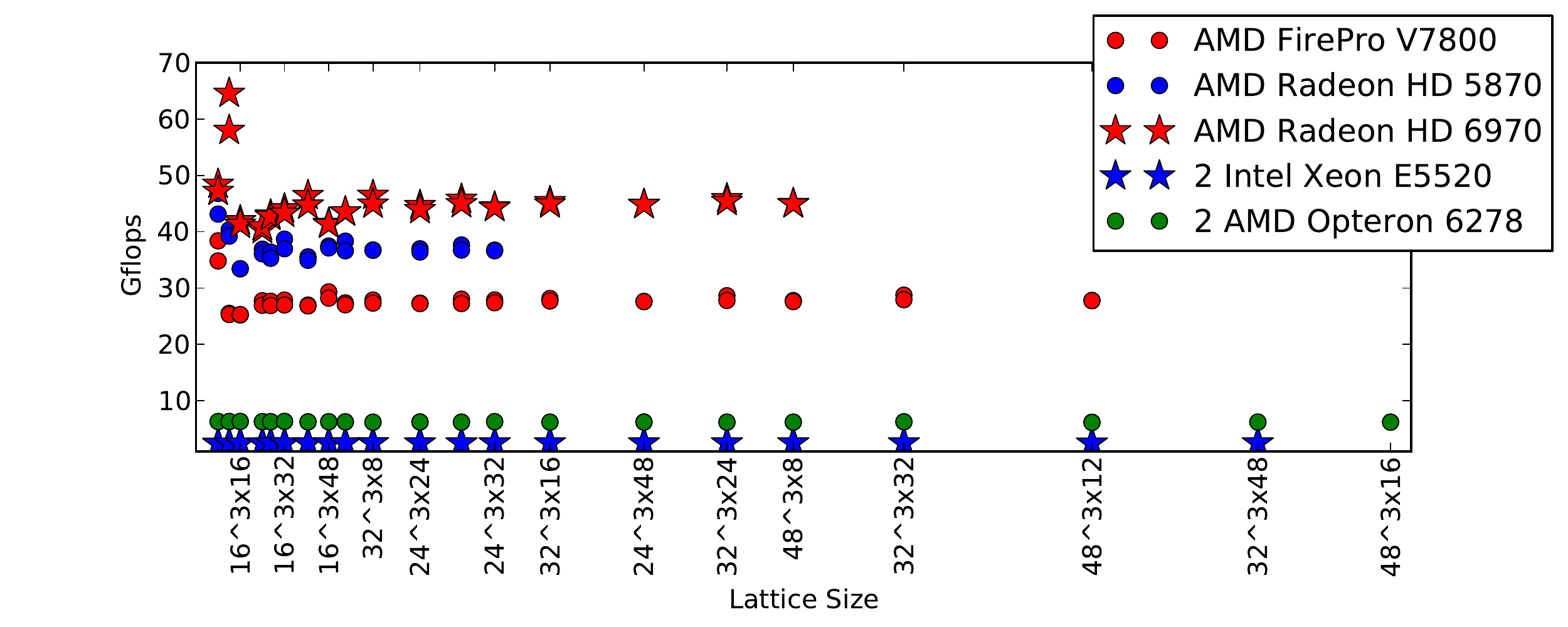}
\includegraphics[width=.8\textwidth]{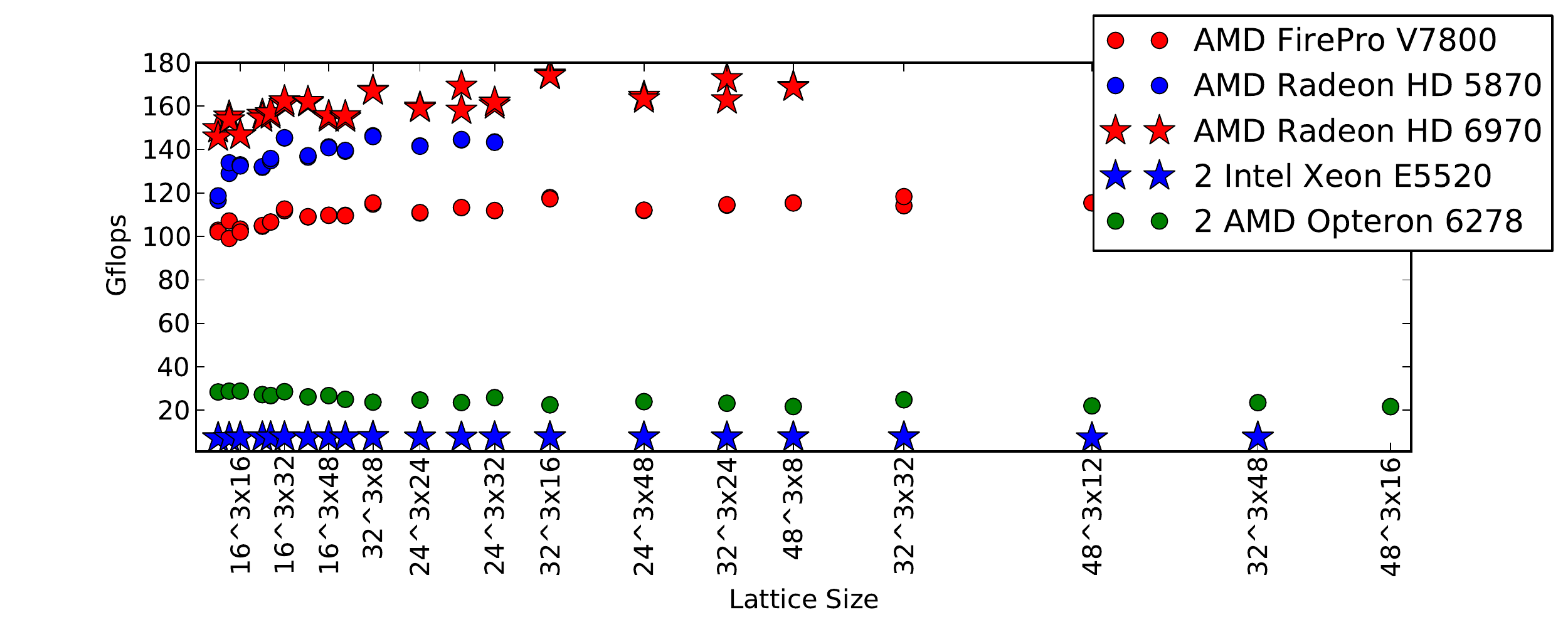}
\caption{Performance of the single precision heatbath and overrelax kernels.}
\label{fig:heatbath-perf}
\end{figure}

First, we present results of our implementation of the heatbath algorithm, that is we concentrate on the heatbath and overrelax kernels. 
Figure \ref{fig:heatbath-perf} shows the performance in GFLOPS achieved in single precision.
While we don't manage to saturate device bandwidth, in both kernels the Cypress and Cayman based GPUs provide at least \unit{30}{GFLOPS} for the heatbath kernel for any lattice size.
The overrelax kernel performs a lot better, here we can achieve at least \unit{100}{GFLOPS} on the GPU, peaking up at about  \unit{160}{GFLOPS} on the AMD Radeon HD 6970.
This behavior can most likely be explained by how the compiler manages with the random number generator, since this is the essential difference between both kernels.
One can also see the different performance of different card generations.
CPU performances are also given, but they are only a factor 1/4 of those on the GPUs.
However, the overrelax kernel again outperforms the heatbath kernel.
The comparison is not entirely fair, as the CPU code did not receive any optimizations besides the proper looping strategy and is not using a SoA access pattern.
Vectorization might be able to close the gap, although one should keep in mind that this is a comparison of two server CPUs to one consumer GPU, where the latter is a much less costly solution.

Not shown is the double precision case, where on the contrary, the performance is of the order of \unit{10}{GFLOPS} and thus poor for both kernels on all GPUs.
Again, the overrelax kernel performs better than the heatbath kernel, however, this time the difference is small.
The massive drop in performance can be understood when looking at the working set sizes of the kernels.
Both kernels require storing of a relatively large amount of data throughout kernel execution, rendering register reuse difficult.
Therefore, in double precision the working set for each thread gets too large to fit into the available registers, leading to massive spilling to scratch registers.
This reduces performance by both, the additional latency when accessing the spilled data, as well as the increased memory bandwidth required to access the spilled data.
Therefore, effective algorithmic density is less than what is given in Table \ref{tab::rw_flops}.
However, this is also specific to heatbath and overrelax kernels, which can be seen in the performances reported below.

Despite performance, the limiting factor for physical studies here is clearly the GPU main memory.
It is limited to \unit{1}{GB} on the AMD Radeon HD 5870 and to \unit{2}{GB} on the other two.
This sets an upper limit for possible lattice sizes.
Additionally, on AMD GPUs prior to the HD 7000 generation there is no official support to use more than half of the GPUs memory from an OpenCL application.
The limit can however be circumvented by setting environment variables as specified by the AMD Knowledge Base\cite{amd-kb-additional-memory}.
Also note that an OpenCL runtime can do a multitude of things when it runs out of physical memory on the GPU.
While current implementations all seem to give an error when running out of memory, from the API specification it is perfectly legal to swap buffers to host memory, which will come with a major performance penalty.

Comparing the heatbath kernel performance with \cite{Cardoso:2011xu}, where performances of a CUDA based heatbath implementation on NVIDIA GeForce GTX 295 and NVIDIA GeForce GTX 580 is given, we observe only half of the reported single precision performances for the heatbath kernel. 
The overrelax kernel performs at around \unit{90}{\%} of the peak value of \cite{Cardoso:2011xu} for the AMD Radeon HD 5870 and a slightly better performance for the AMD Radeon HD 6970 is found.
Note that we do not use memory bandwidth reducing storage techniques for the links as they are used in \cite{Cardoso:2011xu} (\verb+REC12+).
Other than the code in \cite{Cardoso:2011xu}, our implementation can work with any number of lattice points that fit into the GPU memory and is not limited by the number of threads used.
This shows that all the tested GPUs can in principle be used to efficiently run the algorithm.

\subsection{\dslash\ performance}

\begin{figure}[ht]
\centering
\includegraphics[width=.8\textwidth]{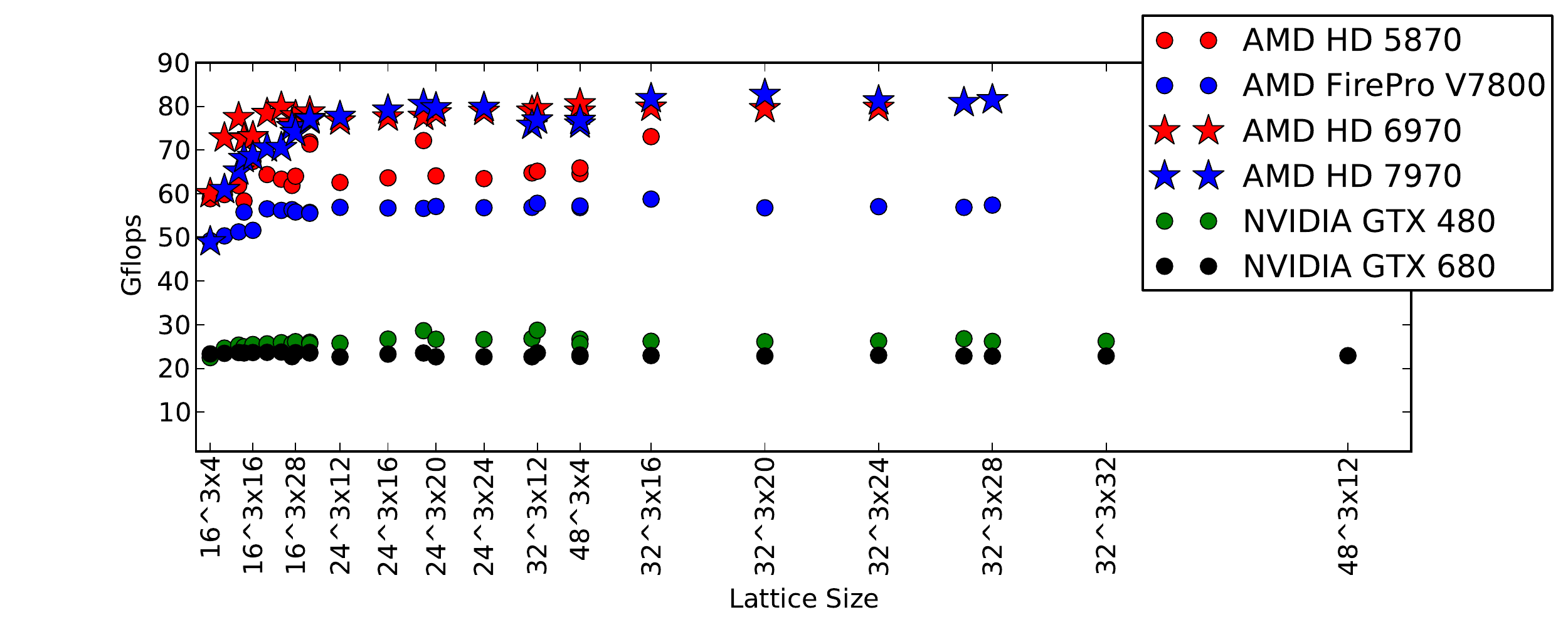}
\includegraphics[width=.8\textwidth]{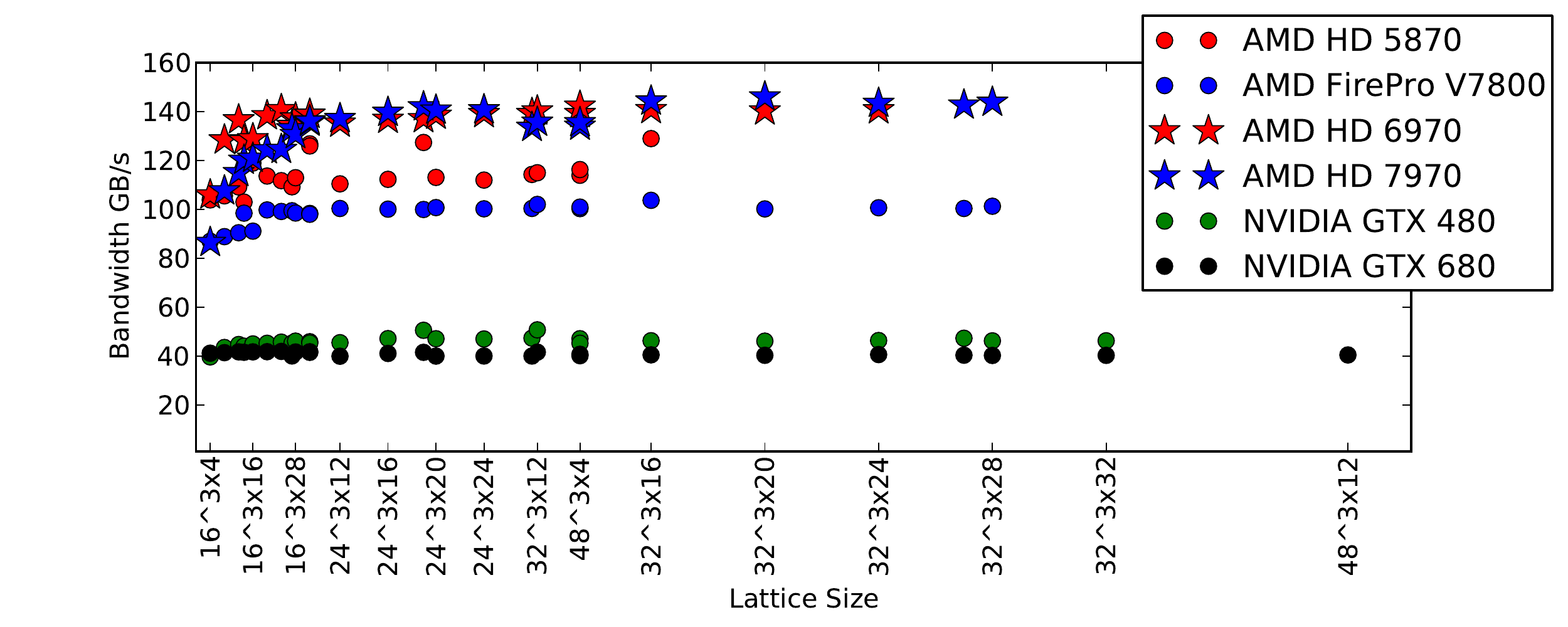}
\caption{Performance and memory bandwidth utilization of the \dslash\ kernel on several GPUs (double precision).}
\label{fig:dslash-dp}
\end{figure}

\begin{figure}[h]
\centering
\includegraphics[width=.8\textwidth]{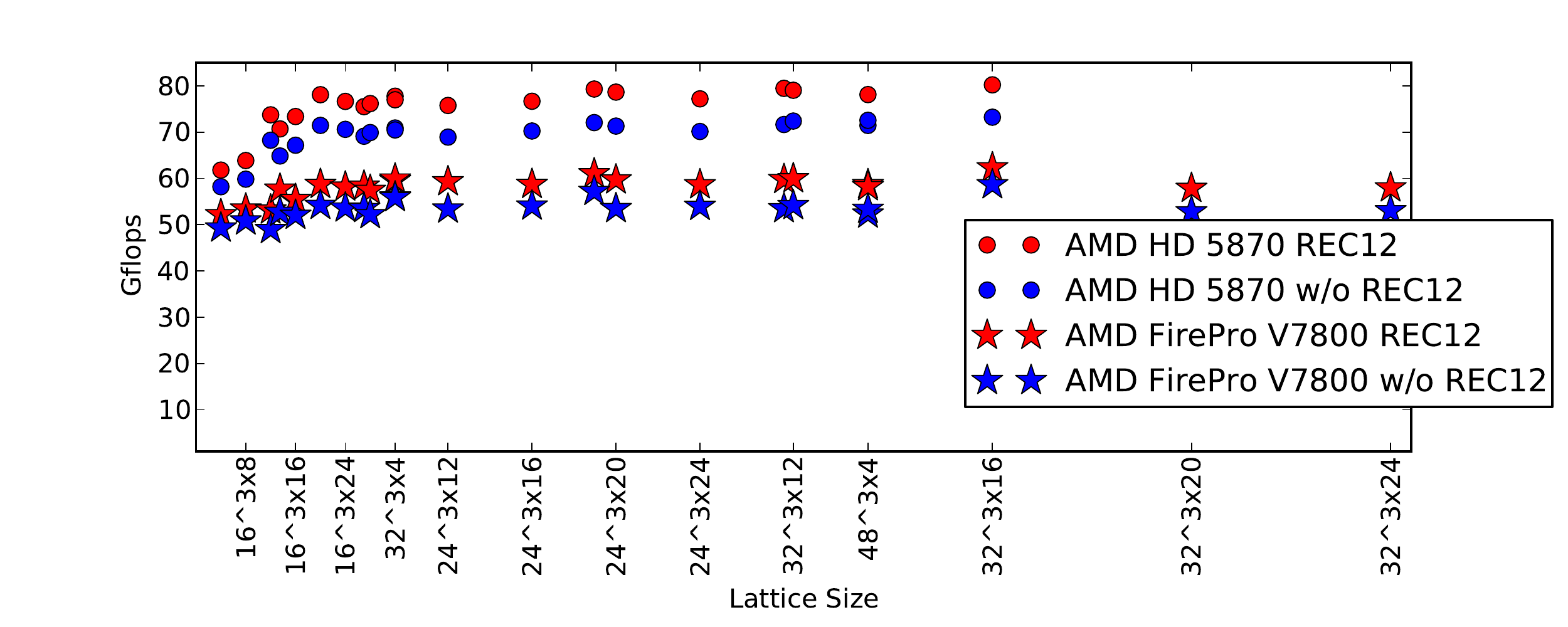}
\caption{Performance of the \dslash\ kernel using REC12 (double precision).}
\label{fig:dslash-dp-rec12}
\end{figure}

For simulations including fermions the inversion of the fermion matrix is essential and thus the performance of evaluating $D \phi$ is crucial. 
Since for twisted mass fermions the diagonal part of $D$ is not dependent on the gauge field, the $\dslash$ performance has to be considered most carefully. 
We measured it on various lattice sizes in double precision, the results can be seen in figure \ref{fig:dslash-dp}.

The slowest of our GPUs, the AMD FirePro V7800 levels at nearly \unit{60}{GFLOPS} and is above \unit{50}{GFLOPS} even for small lattices like \(16^3 \times 4\).
The gamer GPUs AMD Radeon HD 5870 and AMD Radeon HD 6970 scale nicely with their higher peak memory bandwidth, achieving about \unit{60}{GFLOPS} and \unit{80}{GFLOPS}, respectively.
In addition, one can observe better performances with the AMD Radeon HD 5870 for lattices sizes that have a spatial extent that is a multiple of 16, where the kernels performs at about \unit{70}{GFLOPS}.
The performance of the AMD Radeon HD 7970 is not better than that of the AMD Radeon HD 6970, however this GPU was measured using a preview driver.
As that driver does not provide any useful resource usage statistic we did not optimize our kernel for that GPU.
As we cannot ensure proper register usage we simply reused the memory stride optimization found to be optimal on the AMD Radeon HD 5870.

Also shown is the bandwidth utilization, where apart from the unoptimized AMD Radeon HD 7970 at least \unit{70}{\%} of the theoretical peak bandwidth are reached on all AMD GPUs.
On the AMD Radeon HD 6970, when running at \unit{140}{GB/s}, \unit{80}{\%}  of the peak bandwidth are achieved, peaking up to \unit{86}{\%} for lattices like \(16^3 \times 24\).
For the AMD Radeon HD 5870, the utilization peaks at \unit{120}{GB/s} (about \unit{78}{\%} of the theoretical peak bandwidth), which is actually even higher than the about \unit{105}{GB/s} achieved on this GPU in a simple \lstinline{float4} copying kernel as shown in figure \ref{fig:bw_copy}.
This can be explained by the loop-unrolled characteristic of the SoA memory access and the higher read-to-write ratio of the \dslash\ kernel, as only reads can benefit from the cache of the AMD Radeon HD 5870.

Additionally, we give \dslash\ performances obtained on the NVIDIA GPUs GeForce GTX 480 and GeForce GTX 680.
Both cards constantly reach about \unit{25}{GFLOPS} and \unit{40}{GB/s}, respectively.
The latter corresponds to about \unit{20}{\%} of the theoretical peak.
The NVIDIA GPUs show register spilling, which is probably the performance limiting factor.
While the NVIDIA GPUs have a smaller register file than their AMD counterparts, that is only part of the problem.
In CUDA it is possible to have the compiler increase the number of available registers at the cost of fewer threads being able to run on the GPU concurrently.
In addition, it is possible to reconfigure the ratio of L1 cache to shared memory.
A larger L1 cache would probably avoid some of the slowdown due to the spilled registers.
Sadly neither of these features is exposed when using OpenCL on the NVIDIA hardware.
With the current code, the AMD GPUs achieve 3-4 times more performance with the same kernel, but the comparison is not entirely fair, as the AMD GPUs have been the primary development platform.

There are a couple of reported performances in the literature, which are all based on CUDA implementations.
Using the AMD Radeon HD 5870 as our reference value, our $\dslash$ kernel is nearly twice as fast as the one shown in \cite{Clark:2009wm}, even though that kernel uses \verb|REC12| to reduce bandwidth requirements.
One should however keep in mind that the NVIDIA GeForce GTX 280 is one generation older than the AMD Radeon HD 5870.
Our $\dslash$ is also \unit{40}{\%} faster than that shown in \cite{Alexandru:2011ee}.
That performance was measured on a NVIDIA GeForce GTX 480 which, as mentioned above, is as old as the AMD Radeon HD 5870 used by us.

Furthermore, figure  \ref{fig:dslash-dp-rec12} shows the effect of \verb+REC12+ on the \dslash\ performance exemplarily for the AMD Radeon HD 5870 and the AMD FirePro V7800.
This technique reduces the RW load of the kernel by 13\% (see table \ref{tab::mem_req}), and thus should result in a visible speedup.
An increase in performance up to 9\% can indeed be observed and proves again that the performance is bandwidth limited.
Note that we did not include the computational overhead induced by the reconstruction technique in the FLOP count.
Compared to \cite{Clark:2009wm}, where actually a decrease in double precision performance is reported when decreasing the RW load, these results support the good performance picture of our \dslash\ implementation.

\subsection{HMC performance}

\begin {table}
\centering
\begin{tabular}{cccc}
  \hline 
  setup   & A      & B      & C   \\
  \hline
  $a\mu$  & 0.0025 & 0.0035 & 0.1 \\
  $m_\pi$ & 260    & 310    & 520 \\
\end{tabular} 
\caption{Setups used for benchmarking. $\beta = 3.9$ and $\kappa = \kappa_c = 0.160856$ were used throughout to simulate at maximal twist. The value of $m_\pi$ is only approximate. }
\label{tab:setup}
\end{table}

In order to test our HMC program under realistic conditions, we used setups corresponding to one heavy and two lighter pion masses, see table \ref{tab:setup}.
The parameters were chosen according to \cite{Baron:2009wt} in order to simulate at maximal twist.
In each setup, we started from a prior generated gauge configuration and performed 10 HMC steps with $\tau = 1$ (Setup C) and $\tau = 0.1$ (Setup A and B), respectively.
We used separate time scales for the gauge and fermion parts and on each the 2MN integrator with 10 integration steps.
To gain statistics, each run has been carried out $\mathcal{O}(10)$ times.
The OpenCL application was run on the AMD FirePro V7800 and the AMD Radeon HD 6970.
For comparison, we performed the same measurements using the CPU-based application \textit{tmlqcd} \cite{Jansen:2009xp}, running on different numbers of cores on the \loewe\ (AMD Opteron 6172 CPUs).
Statistical errors have been neglected since they were found to be below the percentage limit.

\begin{figure}[h]
	\centering
	\includegraphics[width=.4\textwidth]{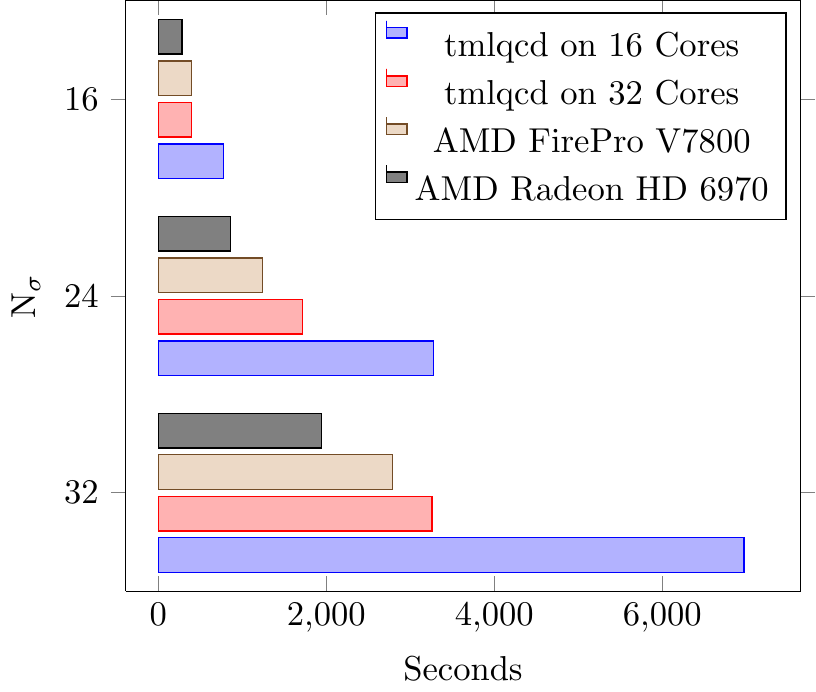}
	\hspace{20mm}
	\includegraphics[width=.4\textwidth]{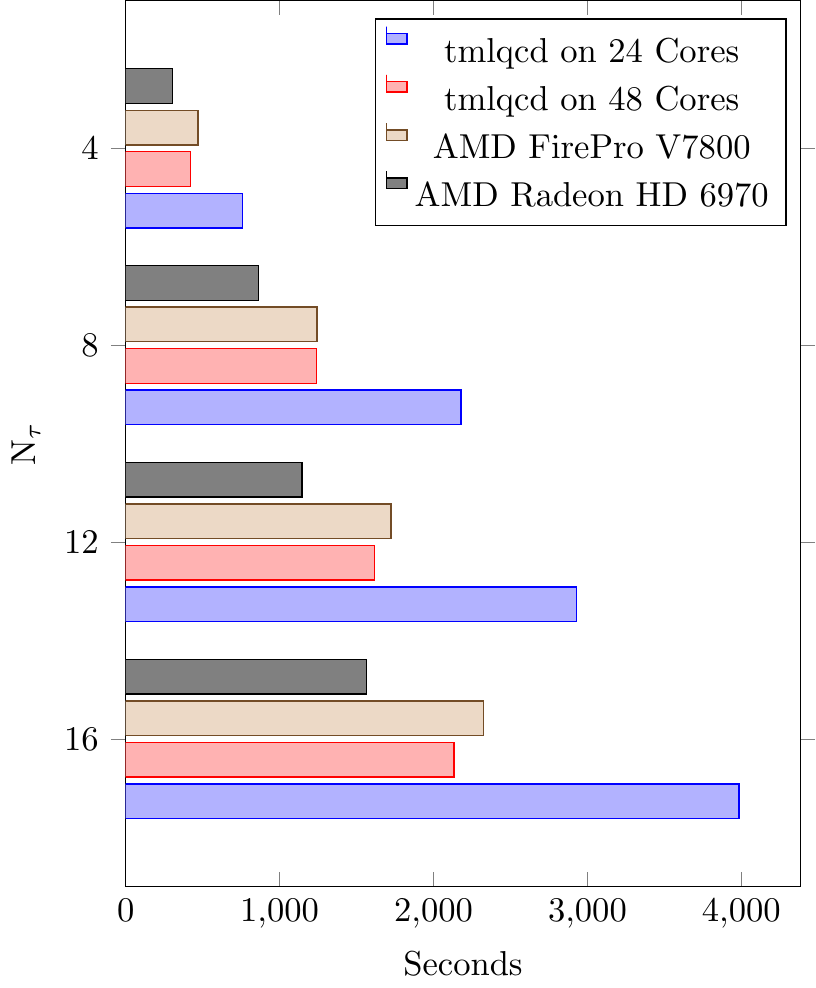}
	\caption{HMC runtimes in seconds for setup C (see table \ref{tab:setup}) for fixed \(\ntau = 8\) and \(\nsigma = 24\), respectively. }
	\label{fig:hmc-results-c}
\end{figure}

The results for setup C are shown in figure \ref{fig:hmc-results-c}, where runs for fixed $\ntau = 8$ and fixed $\nsigma = 24$ have been performed.
The reference \textit{tmlqcd} was executed on one and two whole nodes, which means 16 and 32 and 24 and 48 CPU cores, respectively, for the different lattice sizes, since the lattice extent has to be divisible by the number of cores here.

One can see nice scaling in all reference runs, the additional MPI overhead is only visible for small lattices.
The GPU runtimes are always comparable to or better than the reference values.
Especially the AMD Radeon HD 6970 can achieve a better performance than 32 and 48 CPU cores throughout, respectively.
Considering the $\nsigma = 24$ runs, we measure a speedup of the AMD FirePro V7800 of about 1.7 compared to the 24 core reference runs and a slow-down of 0.9 compared to the 48 core runs.
The AMD Radeon HD 6970 is faster in both comparisons, about 2.5 and 1.3, respectively.
Note that this is a comparison to two and four full CPUs, where each CPU is more expensive then the GPUs used.
We did not perform reference runs on a single CPU core, but we observed a scaling of about 0.75 when running \textit{tmlqcd} on more than one core, which can be explained by the additional MPI overhead.
This scaling yields a speedup of about 30 for the AMD FirePro V7800 and a speedup of about 44 for the  AMD Radeon HD 6970, compared to one AMD Opteron 6172 core.
However, this comparison is a bit academic, since one would practically not use only one core of a multi core CPU.

\begin{figure}
	\centering
	\includegraphics[width=.6\textwidth]{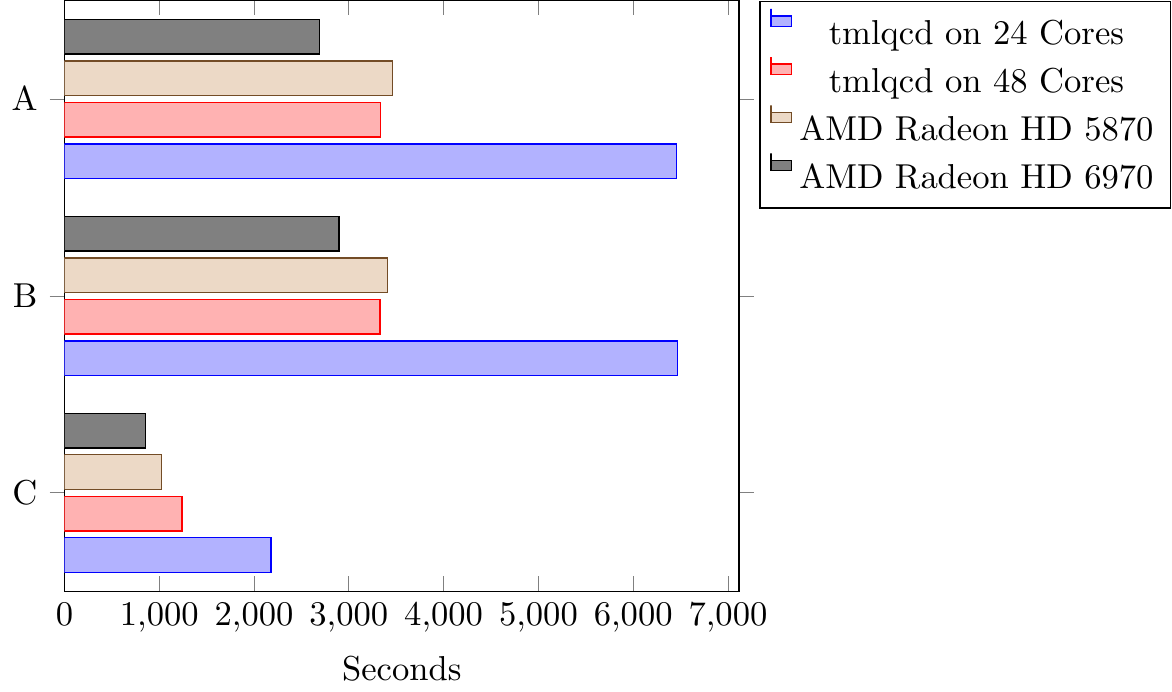}
	\caption{HMC runtimes in seconds for setups A, B, and C (see table \ref{tab:setup}) for a lattice with \(\ntau = 8\) and \(\nsigma = 24\). }
	\label{fig:hmc-results-abc}
\end{figure}

A similar picture emerges also for the lighter pion masses (see figure \ref{fig:hmc-results-abc}), where the runtime scales approximately the same on all systems.
For these setups, the inversion of the fermion matrix takes longer.
Therefore the force calculation, which only achieves about half the bandwidth utilization of the \dslash\ kernel and takes \unit{40}{\%} of the total execution time in setup C, becomes less important for the overall performance.
The limited performance of the force calculation is also caused by register spilling, even though it is not as problematic as in the similar heatbath kernels.

Not displayed are additional runs of the OpenCL code which we did on AMD Opteron 6172 CPUs.
Since no specific optimizations for the CPUs have been carried out, these needed significantly more runtime.
However, switching between both types of devices is smooth, showing that OpenCL code is absolutely versatile.

The comparison of these results to the extensive tests of the CUDA based HMC for the staggered fermion discretization reported in \cite{Bonati:2011dv} is somewhat difficult because of the fundamental differences between the two discretizations. 
Additionally, the authors of \cite{Bonati:2011dv} mainly use single precision within the HMC, whereas we perform double precision calculations throughout.
Although we did not test this, the results obtained for the \dslash\ on NVIDIA cards strongly suggest that our code would currently perform significantly slower on these cards compared to AMD GPUs.
This is the opposite picture to the significantly slower performance of the staggered HMC reported in \cite{Bonati:2011dv}.
Just like our NVIDIA results their AMD results are probably affected by the choice of the primary development platform, resulting in less optimization work in the other.
In addition the staggered fermionic data types should be smaller, reducing the working set and therefore avoiding register spilling.
This should especially benefit the NVIDIA GPUs used, as their register file is smaller than that of the AMD GPU.

\section{Conclusions}
\label{sec:conclusions}

We have presented the first implementation of LQCD using OpenCL and shown it is working on CPUs as well as the two major GPU platforms, NVIDIA and AMD.

In single precision our heatbath implementation is able to achieve a competitive \unit{160}{GFLOPS} on the AMD Radeon HD 6970, while the double precision variant is still work in progress.
Fine tuning the applied memory optimizations, which are currently tuned for double precision codes, and the use of bandwidth-reducing techniques like \verb|REC12| should provide further speedups here.

For the Wilson \dslash~kernel we were able to show excellent performance, utilizing more than \unit{70}{\%} of the available memory bandwidth for all lattice sizes on multiple AMD GPUs outperforming published performances of CUDA based codes.
We also see a positive effect of \verb+REC12+ on the performance.
Extended to a full HMC for twisted-mass Wilson fermions we showed a speedup factor of four of our code running on an AMD Radeon HD 5870 compared to a reference code running on an AMD Opteron 6172.

Further speedups to our HMC will be reached by further optimizing the inverter performance.
One way to reach this will be a mixed precision solver, which up till now we avoided due to the small memory of the AMD Radeon HD 5870.
As proposed in \cite{Philipsen:2011sa} we will also start to use all compute resources on hybrid systems by performing parts of the calculations on the CPU.
This will require the SIMD capabilities of the CPUs, too.
One approach for this could be the redefinition of our base types for each device.
In addition, we are currently investigating possible performance gains by using \verb+REC12+ in the force calculation.
More complex reduction techniques may also be of advantage here.
Finally, to reduce processing time and circumvent memory limitations we will expand our code to run on multiple GPUs.

\section*{Acknowledgments} 
O. P. and C. P. are supported by the German BMBF grant \textit{FAIR theory: the QCD phase diagram at vanishing and finite baryon density}, 06MS9150. M. B., O. P, and C. P. are supported by the Helmholtz International Center for FAIR within the LOEWE program of the State of Hesse. M.B. and C.P. are supported by the GSI Helmholtzzentrum f\"ur Schwerionenforschung. C.P. acknowledges travel support by the Helmholtz Graduate School HIRe for FAIR.

The authors want to thank Lars Zeidlewicz for his participation in the early stages of this project.
Some of the calculations have been performed on LOEWE-CSC. The authors thank the LOEWE-CSC team for all the support.

\bibliographystyle{elsarticle-num}
\bibliography{./literature.bib}

\end{document}